\documentstyle[aps]{revtex}
\input epsf
\input psfig.sty
\input amssym.def
\input amssym.tex

\begin{document}
\draft

\title{Elongation of confined ferrofluid droplets under applied fields\\}
\author{S. Banerjee$^1$, M. Fasnacht$^1$, S. Garoff$^{1,2}$ and M. Widom$^1$}
\address{$^1$Department of Physics \\ 
$^2$ Colloids, Polymers, and Surfaces Program \\
Carnegie Mellon University, Pittsburgh, Pa. 15213}

\date{\today}
\maketitle
\begin{abstract}
Ferrofluids are strongly paramagnetic liquids. We study the behavior of 
ferrofluid droplets confined between two parallel plates with a weak applied 
field parallel to the plates. The droplets elongate under the applied field 
to reduce their demagnetizing energy and reach an equilibrium shape where the 
magnetic forces balance against the surface tension. This elongation 
varies logarithmically with aspect ratio of droplet thickness to its original 
radius, in contrast to the behavior of unconfined  droplets. 
Experimental studies of a ferrofluid/water/surfactant emulsion confirm this 
prediction.

\end{abstract}


\section{introduction}
\label{intro}

Ferrofluids~\cite{rr} are oil- or water-based colloidal suspensions of
permanently magnetized particles. In an applied magnetic field the particles 
align creating a strong paramagnetic response in the ferrofluid. Because they 
are fluids, these suspensions can flow in response to forces. For example, 
ferrofluid droplets elongate parallel to applied 
fields~\cite{d3droplets1,d3droplets2,d3droplets3,ellipsoid,bashtovoi,d2droplets} 
and undergo tip-sharpening
transitions~\cite{sharpen1,sharpen2}. When a ferrofluid droplet is confined
between two plates in a ``thin film'' geometry, surrounded by an 
immiscible fluid, and a field is applied perpendicular to the plates, 
it undergoes field induced bifurcations~\cite{goldstein} leading to 
intricate labyrinthine patterns~\cite{pattern}. Ferrofluid 
emulsions~\cite{bibette} undergo structural 
transitions under an applied field from a randomly dispersed structure of 
the emulsion droplets to droplet chains, columns and worm like 
structures~\cite{liu1,liu2} depending on volume fraction, sample geometry
and the rate of field application.

A droplet of ferrofluid elongates under applied field because of 
the demagnetizing fields of magnetic
poles on the surface of the droplet. Surface poles arise wherever the
droplet magnetization has a component perpendicular 
to the surface. The demagnetizing field that they create opposes the 
magnetization,
creating a demagnetizing energy that depends on the shape of the
droplet. The droplet elongates to reduce its demagnetizing field and
energy. Because elongation increases the surface energy of the
system, an equilibrium shape is reached when the magnetic forces  balance
against the surface tension forces.

The elongation of freely suspended, 3-dimensional droplets
has been well studied~\cite{d3droplets1,d3droplets2,d3droplets3,ellipsoid}. 
The droplets can be assumed to be
ellipsoids for small elongation. The demagnetizing field is thus uniform 
and the elongation (major axis minus minor axis divided by minor axis) is 
found to be proportional to the undeformed 
droplet radius. The case of droplets confined in ``thin film geometry'' 
however, involves two length scales, droplet thickness and its undeformed 
diameter. In the limit of small aspect ratio (droplet thickness divided by its 
undeformed diameter) the demagnetizing fields are stronger near the edges
of the droplet than at its center. We find that the elongation 
divided by droplet thickness in this geometry is proportional to 
the logarithm of the aspect ratio. 

Prior experiments~\cite{d2droplets,bashtovoi} 
have proposed droplet elongation as a tool for measuring surface tension 
between the ferrofluid and the surrounding immiscible fluid. 
We improve on the existing theory~\cite{d2droplets} by incorporating
spatial variation of the demagnetizing field inside the droplet. We perform 
an experiment supporting our predicted logarithmic behavior.

Section~\ref{theory} of this paper presents our theoretical study of the 
elongation of a ferrofluid droplet confined within a thin film. Our principal
result is a predicted logarithmic dependence of elongation on droplet
aspect ratio. We contrast this result with the corresponding elongation of
unconfined droplets. Section~\ref{expt} describes an experiment done with
ferrofluid emulsions that tests our theory. The experiment is in
qualitative agreement with our theoretical prediction, but differs
quantitatively in at least one respect. In section~\ref{discussion} we discuss
a possible explanation of the discrepancy based upon droplet contact angles
with the confining plates.

\section{theory}
\label{theory}

Consider a paramagnetic liquid  droplet confined in a thin film between two 
parallel plates with a gap, $\Delta$, in the ${\hat z}$ direction 
(see figure 1). An immiscible liquid surrounds the droplet. Let the thickness,
$\Delta$, be much smaller than the radius of the undeformed droplet, $r_0$. 
This small aspect ratio
\begin{equation}
p={\Delta \over 2r_0}
\end{equation}
provides the pseudo-two-dimensional character of the problem. If a uniform,
weak, field ${\bf H}_0$ is applied parallel to the plate, the droplet 
magnetizes. The magnetization creates an opposing demagnetizing 
field whose strength depends on the droplet shape. The droplet elongates 
to decrease its magnetic energy, reaching equilibrium when the magnetic 
forces balance against the restoring forces due to surface tension.
In this section we define the elongation of the droplet and calculate
the surface energy, $E_S$, and the magnetic energy, $E_M$, of the droplet as 
a function of its elongation. By minimizing the total energy with respect 
to the elongation we obtain the elongation as a function of ${\bf H}_0$, 
$r_0$, and $\Delta$.

For simplicity assume the elongated droplet has a uniform cross 
section, $\cal C$, independent of $z$. This corresponds to a contact 
angle of $90^\circ$ between the paramagnetic liquid, the surrounding fluid
and the glass plates, and a plate spacing much less than the capillary length
of the two liquids. Thus the droplet has straight edges if viewed from the 
side (see figure 1). The role of 
contact angle will be discussed later in section~\ref{discussion}. 
We write the equation for $\cal C$ in polar coordinates as a generic 
smooth perturbation to a circle,
\begin{equation}
\label{curve}
r=\alpha_1+\alpha_2 \cos 2 \theta.
\end{equation}
We only include a single harmonic, since we expect coefficients for
the higher harmonics to be much smaller than $\alpha_2$ for small
perturbations. The cross section $\cal C$ has semi-major axis $a$, and
semi-minor axis $b$ (see figure 1b), with $\alpha_1=(a+b)/2$ and
$\alpha_2=(a-b)/2$. We define the elongation of the droplet 
\begin{equation}
\label{definition}
\epsilon \equiv {a \over b}-1.
\end{equation}
We assume that the elongation, $\epsilon$, is much less than $1$.
Imposing the constraint that the volume of the droplet ($\Delta$ times
cross-sectional area) remains constant we calculate 
\begin{equation}
\label{alpha}
\alpha_1= {r_0 \over (1+k^2/2)^{1/2}},~~~~\alpha_2={r_0 k
\over (1+k^2/2)^{1/2}},
\end{equation}
where $k=\epsilon / (2+\epsilon)$.

The surface energy is the sum of interfacial areas times surface tensions
between all pairs of the three phases (solid glass, ferrofluid droplet and
immiscible fluid). For the case of uniform cross-section ($90^\circ$ 
contact angle) droplets, the glass-ferrofluid and glass-immiscible fluid
interfacial areas are independent of the shape of ${\cal C}$ due to the
fixed volume constraint. Hence we concern ourselves with the 
droplet-surfactant solution interface, the area of which is $\Delta$ times
the perimeter. The perimeter of cross section $\cal C$ can be calculated 
as a power series in $\epsilon$,
\begin{equation}
\label{circum}
S= 2 \pi r_0 (1+{3 \over 16} \epsilon^2+O(\epsilon^3)).
\end{equation}
As expected, the leading correction to $S$ is second order in $\epsilon$
since the perimeter should increase regardless of the sign of $\epsilon$.
The relevant surface energy of the droplet is 
\begin{equation}
\label{surface-nrg}
E_S=\sigma_{FI} S \Delta 
\end{equation}
where $\sigma_{FI}$ is the surface tension of the
ferrofluid-immiscible fluid interface.

The total magnetic energy of any paramagnetic body under applied field
is~\cite{wfb}
\begin{equation}
\label{magnetic-nrg}
E_M= -{1 \over 2}\int_V d^3{\bf r}~{\bf H}_0 \cdot {\bf M}({\bf r}).
\end{equation}
The magnetization ${\bf M}({\bf r})$ is determined by the self consistent 
equation
\begin{equation}
\label{M}
{\bf M}({\bf r})=\chi({\bf H}_0+{\bf H}_D({\bf r}))
\end{equation}
for linear susceptibility $\chi$, where
\begin{equation}
\label{demag-field}
{\bf H}_D({\bf r})= \int_{S} d^2 {\bf r'}~ 
({\bf M}({\bf r'}) \cdot  {\bf \hat{n}}({\bf r'}))
{ {\bf r-r'} \over |{\bf r-r'}|^3}
+\int_{V} d^3 {\bf r'}~
(\nabla \cdot{\bf M}({\bf r'}))
{{\bf r-r'} \over |{\bf r-r'}|^3}
\end{equation}
is the demagnetizing field due to the magnetization ${\bf M}({\bf
r})$, with ${\bf {\hat n}}({\bf r'})$ being the outward normal at any
point on the surface. The surface integral gives the demagnetizing
field due to the surface poles which appear wherever the magnetization
has a component normal to the surface. The volume integral gives the 
contribution to the demagnetizing field due to volume charges which 
appear at points where the magnetization has non-zero divergence.

To calculate the magnetic energy we expand ${\bf M}$ and ${\bf H}_D$ 
in power series in the susceptibility $\chi$,
\begin{equation}
\label{mag}
{\bf M}({\bf r})={\bf M}^{(1)}({\bf r})+{\bf M}^{(2)}({\bf r})
+{\bf M}^{(3)}({\bf r})+...
\end{equation}
\begin{equation}
\label{demag}
{\bf H}_D({\bf r})={\bf H}_D^{(1)}({\bf r})+{\bf H}_D^{(2)}({\bf r})+
{\bf H}_D^{(3)}({\bf r})+... ,
\end{equation}
where ${\bf M}^{(n)}({\bf r})$ and ${\bf H}_D^{(n)}({\bf r})$ are proportional 
to $\chi^n$. Equating terms in~(\ref{M}) of equal order in $\chi$ we get
\begin{equation}
\label{first-order}
{\bf M}^{(1)}({\bf r})=\chi {\bf H}_0
\end{equation}
and
\begin{equation}
\label{higher-order}
{\bf M}^{(n+1)}({\bf r})=\chi {\bf H}_D^{(n)}({\bf r}).
\end{equation}
Note that ${\bf M}^{(1)}({\bf r})$ is independent of ${\bf r}$ because
the applied field is uniform whereas ${\bf M}^{(n)}({\bf r})$ may depend on
$({\bf r})$ for $n>1$ because ${\bf H}_D({\bf r})$ may be non-uniform. To 
second order in $\chi$ we write the magnetic energy of the droplet 
in~(\ref{magnetic-nrg}) as
\begin{equation}
\label{mag-nrg2}
E_M=-{1 \over 2}\chi{H}_0^2 V -{1 \over 2}\int_V d^3{\bf r}
~{\bf M}^{(1)}\cdot {\bf H}_D^{(1)}.
\end{equation}

The first term in equation~(\ref{mag-nrg2}) for the magnetic energy is
independent of the shape of the droplet and hence unimportant for our
consideration. The second term in the energy is the demagnetizing
energy $E_D$ due to a uniform magnetization ${\bf M}^{(1)}=\chi{\bf
H}_0$. Because ${\bf M}^{(1)}$ is uniform there are no volume charges,
and the surface poles appear only along the droplet-immiscible
fluid interface, to first order in $\chi$.  Rewrite the second term
in~(\ref{mag-nrg2}) as an energy due to the induced surface charges
along the curved surface of the droplet 
\begin{equation}
\label{demag-nrg}
E_D={1 \over 2} \chi^2  \int_0^\Delta dz \int_0^\Delta dz'
\oint ds \oint ds' {({\hat {\bf n}} \cdot {\bf H}_0)
({\hat {\bf n'}} \cdot {\bf H}_0) \over |{\bf r}-{\bf r'}|}.
\end{equation}
Here $ds$ and  $ds'$ are infinitesimal arc-lengths along the contour of the 
droplet ${\cal C}$, and ${\hat {\bf n}}$ and ${\hat {\bf n'}}$ are the outward 
normals to the curved surface of the droplet at points $(s,z)$ and $(s',z')$ 
respectively.

Write $|{\bf r}-{\bf r'}|= {\sqrt {R^2+(z-z')^2}}$, where 
$R$ is the in-plane distance between points at 
positions $s$ and $s'$ on $\cal C$. Integrating over $z$ and $z'$
in~(\ref{demag-nrg})  yields~\cite{goldstein}
\begin{equation}
\label{demag-nrg2}
E_D=  \chi^2 \Delta \oint ds \oint ds' ({\hat {\bf n}} \cdot {\bf H}_0)
({\hat {\bf n}}' \cdot {\bf H}_0) \Phi(R/ \Delta)
\end{equation}
where
\begin{equation}
\Phi (R/\Delta)= R/\Delta - {\sqrt {1+(R/\Delta)^2}} + 
\ln {\Big [}(R/\Delta)/({\sqrt {1+(R/\Delta)^2}}-1){\Big ]}.
\end{equation}
Using equation~(\ref{curve}) for $\cal C$ we calculate the
demagnetizing energy in~(\ref{demag-nrg2}) as a series expansion in
$\epsilon$ and the aspect ratio $p=\Delta/2r_0$
\begin{equation}
\label{expansion}
E_D=  \chi^2 H_0^2 V {\Big\{}{2p} \ln {B \over p}- 3\epsilon p  
\ln { C \over p}+ \cdot \cdot \cdot {\Big \}}
\end{equation}
where $V=\pi r_0^2 \Delta$ is the volume of the droplet, and $B=4
e^{-1/2}$ and $C=4 e^{-5/6}$ are geometrical constants. The term in
the brackets can be identified as $2 \pi$ times the demagnetizing 
factor~\cite{wfb}
of the droplet along the direction of applied field. Additional terms
in the series in equation~(\ref{expansion}) are of higher order in
$\epsilon$ or in $p$. For small elongation and large aspect ratio
we may neglect these higher order terms.

Minimizing the total energy $E=E_S+E_M$ with respect to $\epsilon$ gives
\begin{equation}
\label{elongation}
\epsilon= {\chi^2 H_0^2 \Delta \over \sigma_{FI}}\ln {C \over p}.
\end{equation}
Corrections to this result are higher order in aspect ratio $p$ or higher 
order in $\epsilon$ itself. Interestingly, the elongation depends only 
logarithmically on the undeformed radius $r_0$, and has a much stronger 
dependence on the thickness, $\Delta$, of the droplet.  This result differs
from an earlier theory~\cite{d2droplets} which omits the logarithm because
it assumes that the demagnetizing field is uniform inside the droplet. 

In the case of unconfined, nearly ellipsoidal 
droplets~\cite{d3droplets3,ellipsoid},
the demagnetizing field is quite uniform inside the droplet. The demagnetizing
energy is therefore proportional to the volume ($(4/3)\pi r_0^3$) of the 
droplet according to equation~(\ref{mag-nrg2}). The surface energy is 
proportional to the area ($4 \pi r_0^2$) and the elongation is thus 
proportional to $r_0$. In the case of thin film geometry, however, the 
demagnetizing field is very non-uniform. For distances much less than $\Delta$ 
near the droplet edge, the component of the demagnetizing field is of order 
$M$, since the edge acts like an infinite sheet of charge in the first 
approximation. For distances much greater than $\Delta$ the demagnetizing 
field  is of order $M \Delta/r$ since the edge acts as a line charge in 
this case. The contribution to the integral for the 
demagnetizing energy in equation~(\ref{mag-nrg2})  mainly comes from the bulk
of the droplet and goes like $r_0 \Delta^2 \ln (r_0/\Delta)$. The 
surface energy is proportional to $2 \pi r_0 \Delta$ and the 
elongation is therefore proportional to $\Delta \ln (r_0/\Delta)$. The 
logarithmic variation of elongation with the aspect ratio is thus a 
signature of the non-uniform nature of the demagnetizing field inside the 
droplet.

\section{experiment}
\label{expt}

\subsection{Setup}
\paragraph{Sample Preparation and Structure}

Our sample consisted of a ferrofluid/aqueous solution emulsion confined 
between two glass plates. The oil-based ferrofluid used was EMG 905 made 
by Ferrofluidics. To reduce the surface tension between the ferrofluid and
the immiscible aqueous external  phase, we incorporated surfactants in 
the aqueous phase. A solution of a commercial detergent made the best 
emulsions while solutions
with other pure anionic surfactants either showed hardly any elongation 
of the ferrofluid droplets under applied field or produced droplets 
without sharp boundaries with the aqueous phase. In contrast, our stable, well 
behaved emulsions allowed us to probe and confirm the fundamental aspects 
of our model.

To prepare the emulsions,  a single drop of ferrofluid ($\sim 0.1$ ml) 
was added to $10$ ml of surfactant solution which was a $12$ times dilution
of the commercial detergent. The liquid was shaken (by hand) to prepare the 
emulsion, creating ferrofluid droplets with diameters varying from 
$\sim 5-200~\mu$m. A small amount of this emulsion was then put between 
two glass plates which were circular, about $2$ cm  in diameter and  $4$ mm 
in thickness. These plates were cleaned using soap and alcohol and then rinsed 
with ROPure water. We also tried acid cleaning of the glass plates, however it 
did not result in any noticeable change in the quality of the sample.

We used a rectangular spacer made of mylar foil to separate the plates and
prevent the emulsion from leaking out from the edges of the plates. The mylar 
foil extended to to the edges of the glass plates and had a rectangular
 hole in the center into
which the emulsion was inserted. The thickness of a single mylar spacer was 
measured to be $6.54\pm 0.06~\mu$m. The experiment was performed  with one 
and two spacers to ensure small aspect ratio.

For the cell assembly, the mylar spacers were placed on the first plate 
and a drop of the emulsion was put in the center of the plate. The second 
plate was placed on top and the two plates were  clamped together using 
a pair of brass rings. The rings were tightened by a set of $4$ equally 
spaced screws. We measured the thickness variation across the sample by 
making a ``dry'' sample (without the emulsion) and counting resulting 
white light interference fringes. Although the thickness of mylar spacers 
was measured to an accuracy of $1$ percent, the thickness variation across 
the sample was found to be $10 \%$ resulting from  the stresses 
due to clamping and possible entrapment of dust in the cell.

\paragraph{Apparatus}
A schematic diagram of the experimental setup is shown in figure 2. 
We put the sample at the center of a pair of Helmholtz coils 
to insure a homogeneous magnetic field. The field measured 
close to the sample using a Hall probe showed a variation of less than 
$4\%$ across the sample. The sample was set up horizontally to prevent 
gravitational settling of the ferrofluid droplets. Horizontal alignment
was achieved using a bubble level.

The sample was illuminated from below using a diffused light source and
observed from above using a tele-microscope. The 
tele-microscope was connected to a CCD camera  and the image from it was 
fed into a video recorder and recorded on video tape. Images 
from the recording were later processed using {\it NIH Image}. We calibrated 
the optical system using a measuring reticule aligned along the two 
orthogonal directions of the CCD array. Figure 3 
shows a low magnification view of a typical sample. The ferrofluid droplets 
appear much darker in the image than the surfactant solution around them.

\paragraph{Experimental Procedure and Image Analysis}
During the experiment the applied field was incremented every few seconds.
We found the response of the droplets to the field to be  nearly 
instantaneous and the shape of the droplets remained constant at constant 
field. Experiments with decreasing field strength showed no hysteresis in 
droplet shape. While droplet elongations were observed to be small we 
incremented the field in steps of about $1$ Gauss, and increased the increments
up to about $5$ Gauss as the elongation increased. Droplet elongations appeared
to vary smoothly with applied fields over the entire range from $0$ to $50$
Gauss.

During each experiment the droplets were observed on a video monitor and 
recorded on tape. After grabbing images of distorted droplets, we used a 
cut-off in pixel gray scale level to identify the droplet edge. The 
semi-major axis  $(a)$ and the semi-minor axis $(b)$ were directly read off 
the image using {\it NIH Image}. At zero field measured elongations were small 
(RMS magnitude around $0.003$) and in random directions. These minor 
perturbations from a circular shape were likely due to microscopic distortion
of the contact line pinned on weak surface heterogeneities. The ``observed 
radius'' $r_0$ was calculated as the average of the two semi-axes at zero 
field and the elongation at each field value was calculated using data 
analysis software.

\paragraph{Results}
For each of the $48$ droplets studied  we plotted 
elongation  $\epsilon$, versus the square of the applied field, ${\bf H}_0$. 
Figure $4$ shows typical plots. The elongation is  proportional to the 
square of the applied field for small applied fields as predicted. 
Saturation effects, although small, can be seen at higher values of the field. 
The plot of elongation, for each droplet was  fitted to
\begin{equation}
\label{fit}
{\epsilon \over \Delta}=k_0+k_1 H_0^2 + k_2 H_0^4.
\end{equation}
We included terms only up to order $H_0^4$ because the saturation effects
were observed to be small. We include $k_0$ to allow for the observed small 
elongations at zero field. 

The coefficients $k_1$ of each droplet were then plotted versus the inverse
of the aspect ratio $1/p=2 r_0 /\Delta$  
on a semi-log plot (see figure 5). The theory predicts a slope of $\chi^2 /
\sigma_{FI}$ and an intercept of $1/C$ on the horizontal axis with $C=1.74$.
The data points in figure 5(a) fall on a straight line as predicted 
by the theory. Also, as predicted by the theory, the data points for
two different droplet thicknesses overlay each other.
There is substantial scatter in the data, but the deviations from a 
straight line are random and consistent with the error bars.  The chief 
source of uncertainty was the $10 \%$ uncertainty in thickness due to the 
variation observed across the sample.  Figure 5(b) displays the deviation 
of $k_1$ from the best fit normalized by the uncertainty. The uncertainties  
in measuring $\epsilon, r_0$ and ${\bf H}_0$ were found to be negligible in 
comparison. 

Dividing the susceptibility $\chi=1.9$ for the ferrofluid 
used~\cite{ferrofluidics} by the slope 
= $0.119\pm0.004$~cm/dyne obtained from the fitted line we get 
$\sigma_{FI}=30.4 \pm 1.1$ dynes/cm,
typical of oil-water surface tensions. 
From the fitted line we also get $C=0.35\pm 0.08$, differing substantially 
from our theoretically predicted value of $1.74$. It may be 
possible to explain this discrepancy by considering the effect of the 
contact angle of the ferrofluid-immiscible fluid interface with the glass 
plates.  In the discussion section below we explore the qualitative 
effect of the contact angle.

In figure 6(a) we plot $\epsilon /\Delta$ versus $2r_0/\Delta$ on a linear
scale. If the demagnetizing field inside the droplet was uniform like in 
the case of unconfined droplets,
the plot would be a straight line. However, the plot is clearly not a straight
line and the deviations from the best fitted straight line are systematic (see
figure 6(b)).
This further supports our theoretical result that the demagnetizing field
inside a confined droplet is non-uniform and the elongation divided by 
thickness is proportional to the logarithm of the aspect ratio.

\section{Discussion}
\label{discussion}

The results discussed in section~\ref{expt} agree with
our theoretical prediction~(\ref{elongation}) of logarithmic variation
of $\epsilon/\Delta$ with a proportionality constant of $\chi^2/\sigma_{FI}$. 
However, our theoretical value for $C$ is $4e^{-5/6}=1.74$ whereas the 
experimentally measured value for $C$ is $0.35 \pm 0.08$. 

One possible explanation for the discrepancy in the value of $C$ is
that the ferrofluid/glass contact angle is not $90^\circ$ and consequently
the cross-section of the droplet is not uniform. Our calculations
are for uniform droplet cross-section, which corresponds to a contact
angle $\beta=90^\circ$ between the glass plate and liquid droplet.
The experiment, however, was performed with an oil-based ferrofluid in
a surfactant solution for which the oil-glass contact angle $\beta <
90^\circ$ (see figure 7). A contact angle of other than $90^\circ$ 
will affect the elongation in two ways: by changing interfacial areas
to alter the functional form of $E_S$ and  by redistributing the magnetic
surface poles to alter the functional form of $E_M$. We consider these
two effects in turn. First, however, we must address an ambiguity in
the definition of aspect ratio and elongation which results from the
non-uniformity of droplet cross-section.

Our experiment observes the profile of the largest cross-section of
the droplet. For a circular droplet with $\beta<90^\circ$ this is the 
 radius $r_1$ defined as the radius at mid-gap as shown in figure 7.
For an elongated droplet we measure the semi-major and -minor
axes $a_1$ and $b_1$ and, through equation~(\ref{definition}), the
elongation $\epsilon_1$. We also define $r_2$, $a_2$, $b_2$ and
$\epsilon_2$ associated with the ferrofluid-immiscible fluid-glass
plate contact line (see figure 7). Since $\Delta$ is much less than the
capillary length of the ferrofluid/immiscible fluid, to a good 
approximation~\cite{contact} the profile of
the droplet will be an arc of a circle, so the difference between
$r_1$ and $r_2$ is of order $\Delta$, and likewise for the semi-major
and -minor axes. The difference $\epsilon_1-\epsilon_2$ is of order
$\Delta/r_1$ relative to the elongation.  Recall that our
result~(\ref{elongation}) for the elongation is only the lowest order
term in a series expansion in the aspect ratio. Thus the distinction
between $r_1$ and $r_2$, and between $\epsilon_1$ and $\epsilon_2$,
does not alter our result at the lowest order in aspect ratio.

When the contact angle differs from $90^\circ$, the cross section of
the droplet depends on $z$. Consequently, the contact areas of the
glass plates with the droplet and with the surfactant solution may
vary as the droplet elongates. All the three interfacial areas must be
taken into account to calculate the surface energy. The total surface
energy is
\begin{equation}
E_S= \sigma_{FI} A_{\cal C}+2\sigma_{FG}A_G+2\sigma_{IG}(A-A_G),
\end{equation}
where the three surface tensions between ferrofluid-immiscible fluid, 
ferrofluid-glass, and surfactant solution-glass, are denoted by 
$\sigma_{FI},\sigma_{FG}$, and $\sigma_{IG}$ respectively, $A_C$ and $A_G$
are defined below and the total area
of the sample is denoted by $A$. The factors of $2$ in the second and
third terms of the surface energy account for the two glass surfaces. 

The area of the droplet-surfactant solution  interface $A_{\cal C}$ is 
given approximately by the circumference of $\cal C$ multiplied by the 
arc length of the bulge
\begin{equation}
\label{AC}
A_C=2 \pi r_1 (1+{3\over 16}\epsilon^2)
\Delta {(\pi/2-\beta) \over {\cos \beta}}.
\end{equation}
We use $r_1$ here to calculate the circumference of the droplet because is
the radius observed during the experiment. To first order in the aspect ratio,
using $r_1$ or $r_2$ in equation~(\ref{AC}) yields the same result.

The droplet's contact area with the glass plates must be adjusted
to maintain a constant total volume of ferrofluid as the droplet elongates. 
We approximate the volume  of the bulging region by the circumference of 
$\cal C$ multiplied by the projected area of the bulge. The contact area
$A_G$ must be adjusted so that $A_G \Delta$ changes by the negative of the
change in volume of the bulge. Thus we write
\begin{equation}
A_G=2\pi r_2^2 {\bigg [}1- {3 \over 32} {\bigg \{}{(\pi/2-\beta)\over 
{\cos^2 \beta}}-{\tan \beta}{\bigg \}} {\Delta \over r_2} 
\epsilon^2{\bigg ]}.
\end{equation}
Using $r_2$ instead of $r_1$ makes the above result exact for zero elongation.
The area of ferrofluid in contact with the glass plates decreases with 
elongation for an acute contact angle because the volume of the fluid 
contained in the outward bulge of the droplet increases and therefore the 
fluid contained in the bulk of the droplet decreases. For obtuse contact 
angles exactly the opposite happens for similar reasons.

To understand how the contact angle affects the magnetic energy,
consider the work done by the magnetic field as we change the contact
angle from $90^\circ$ to $\beta$ while keeping the volume of the
droplet constant.  This work, divided by the circumference ,
must be independent of $r_1$ in the limit $r_1$ going to infinity,
since the magnetic field near the surface of the droplet will not depend
on $r_1$ in the large $r_1$ limit. The work done by the magnetic field
is the difference in energy between the straight edge droplet with
contact angle of $90^\circ$ and the bulging droplet with a contact
angle of $\beta$. The demagnetizing energy of the bulging droplet must
therefore have the same dependence on $\ln ({r_1/\Delta})$ as the
straight edge droplet or the difference in the demagnetizing
energies divided by the circumference will be proportional to $\ln
({r_1/\Delta})$ and will blow up in the large $r_1$ limit. Hence, the
demagnetizing energy for the bulging droplet must be identical to
equation~(\ref{expansion}) but with different values $\tilde{B}$ and
$\tilde{C}$ replacing the constants $B$ and $C$.

As the droplet bulges inward or outward the charges on the surface get
distributed over a larger area, decreasing the demagnetizing energy.
The constant $\tilde{B}$ therefore has a smaller value for a bulging
(inward or outward) droplet than $B$, the value for a straight-edged
droplet. However, since the demagnetizing energy is always positive,
smaller demagnetizing energy ($\tilde{B} < B$) implies a weaker
dependence of demagnetizing energy on elongation. Thus, we expect the
value of $\tilde{C}$ to be smaller for a bulging droplet than the
value $C$ for a straight-edged droplet.

Finally, consider how the contact angle dependence of surface and magnetic
energies affect the elongation calculated in equation~(\ref{elongation})
for the case $\beta=90^{\circ}$. The $\epsilon$ dependence of the
surface energy remains quadratic, but the coefficient now depends upon
a linear combination of the three surface tensions $\sigma_{FI}$,
$\sigma_{FG}$ and $\sigma_{IG}$. This combination will replace
$\sigma_{FI}$ in equation~(\ref{elongation}). The functional form of
the magnetic energy remains unchanged, but the values of $B$ and $C$
depend on contact angle. Thus the smaller value $\tilde{C}$ replaces $C$ in
equation~(\ref{elongation}). For $\beta \ne 90^\circ$ the experiment
cannot be used to determine $\sigma_{FS}$ unless $\sigma_{FG}$ and 
$\sigma_{IG}$ are known. Since the $\beta$ in general is not  
$90^\circ$, it is only possible to  measure the effective surface 
tension during elongation, and not  $\sigma_{FI}$ itself.

\section{Conclusions}
We study the elongation ferrofluid droplets, confined in thin film geometry,
under weak applied field. Our theoretical calculations predict the elongation 
of a droplet depends logarithmically on aspect ratio. This behavior 
contrasts with the case of unconfined 3-dimensional droplets where elongation
is directly proportional to undeformed droplet radius. We measured the 
elongation of ferrofluid droplets in an experiment performed on ferrofluid 
droplets in a ferrofluid/water/surfactant emulsion. The results of our 
experiment agree with the functional form our theoretical prediction, 
however the experimentally measured value of $C$ differs from the 
predicted value.We suggest the droplet contact angle with the confining 
plates as a source of this discrepancy.

\acknowledgements
We acknowledge partial support for this research under NSF grant DMR-9732567.

\newpage

\begin{figure}[tb]
\epsfxsize=6in \epsfbox{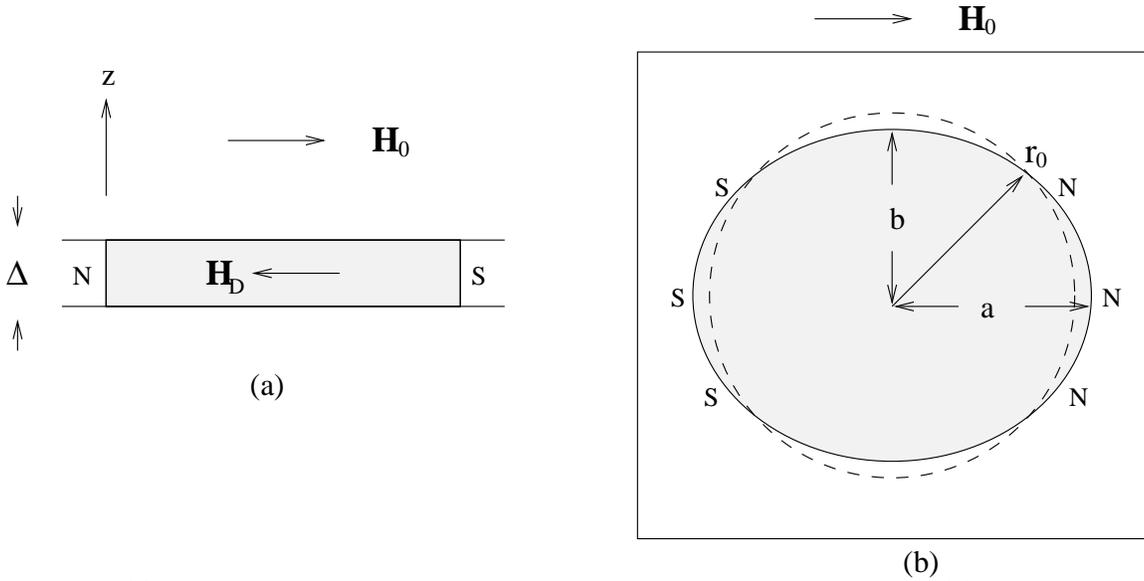}
\caption{(a) A side view of a ferrofluid droplet confined between 
two glass plates.
(b) A top view of a ferrofluid droplet elongating under applied field. The 
dashed line shows the undeformed droplet. N and S indicate the north and 
south magnetic poles. ${\bf H}_D$ is the demagnetizing field.}
\label{f1}
\end{figure}

\begin{figure}[tb]
\epsfxsize=3in \epsfbox{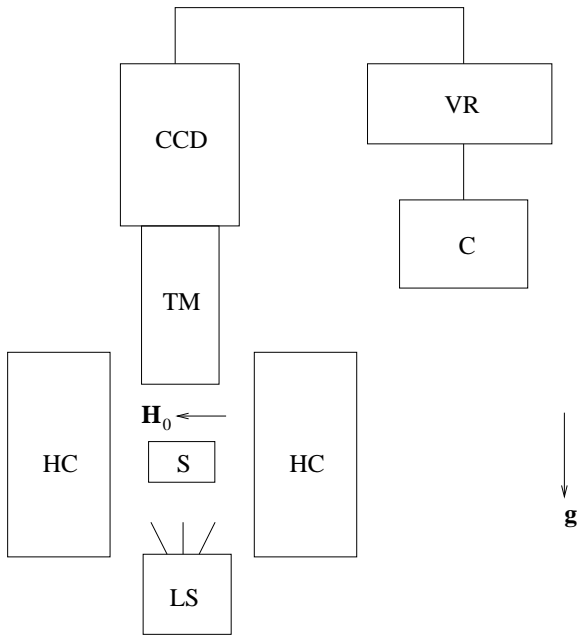}
\vspace{0.5cm}
\caption{A schematic diagram of the experimental setup.  LS=Light Source, 
HC=Helmholtz Coil, S=Sample, TM=Tele-Microscope, CCD=CCD Camera, VR=Video 
Recorder, C=Computer}
\label{f2}
\end{figure}

\begin{figure}[tb]
\epsfxsize=3in \epsfbox{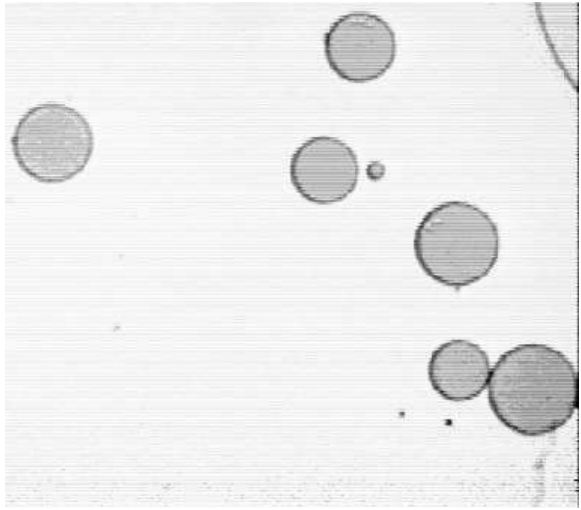}
\caption{A small magnification view of the sample showing ferrofluid droplets
in emulsion.}
\label{f3}
\end{figure}

\begin{figure}[tb]
\psfig{figure=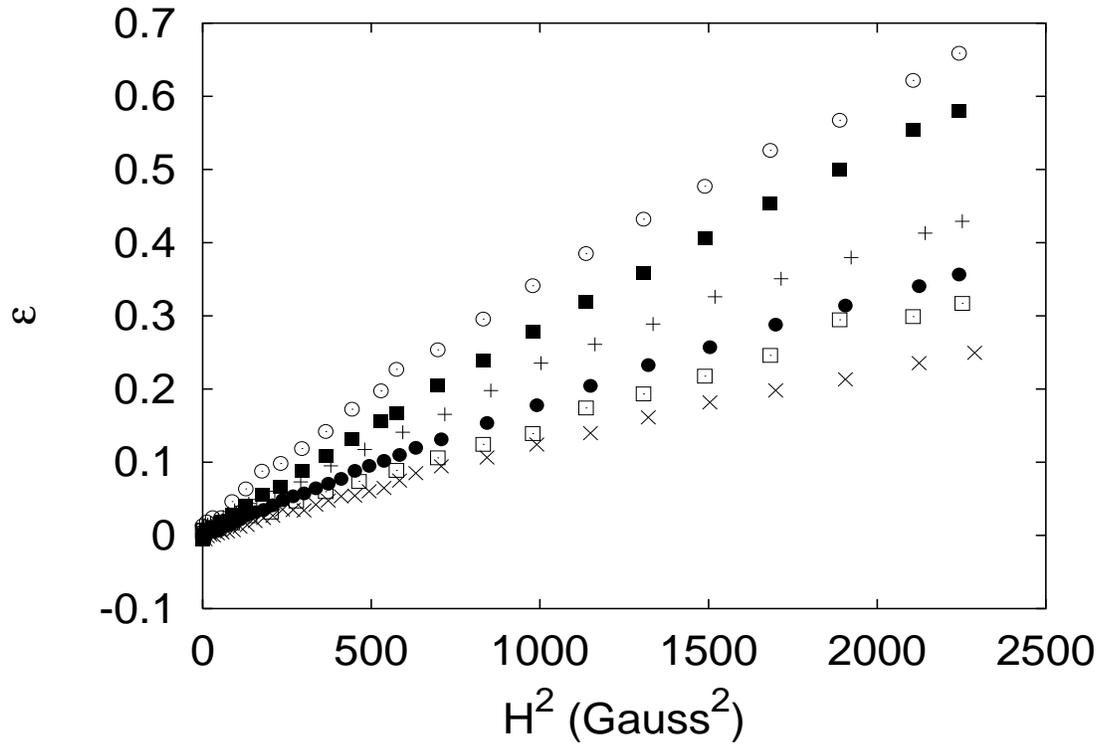,angle=-90,height=4.0in,width=6.0in}
\vspace{0.5cm}
\caption{The plot of elongation vs $H^2$ for droplets with different radii
 and two different thicknesses. The error bars are smaller 
than the size of the symbols on the plot. The symbols and radii for droplets with 
thickness spacing of $\Delta=6.5~\mu m$ are $(\times,~96.5~\mu m)$, 
$({\Huge \bullet},~112.0~\mu m)$ 
and  $(+,~216.5~\mu m)$. The symbols and radii for 
droplets with  thickness spacing of $\Delta=13.1~\mu m$ are 
$(\boxdot,~50.0~\mu m)$, $(\blacksquare,~98.0~\mu m)$ and $(\odot,~177.0~\mu m)$.}
\label{f4}
\end{figure}

\begin{figure}[tb]
\psfig{figure=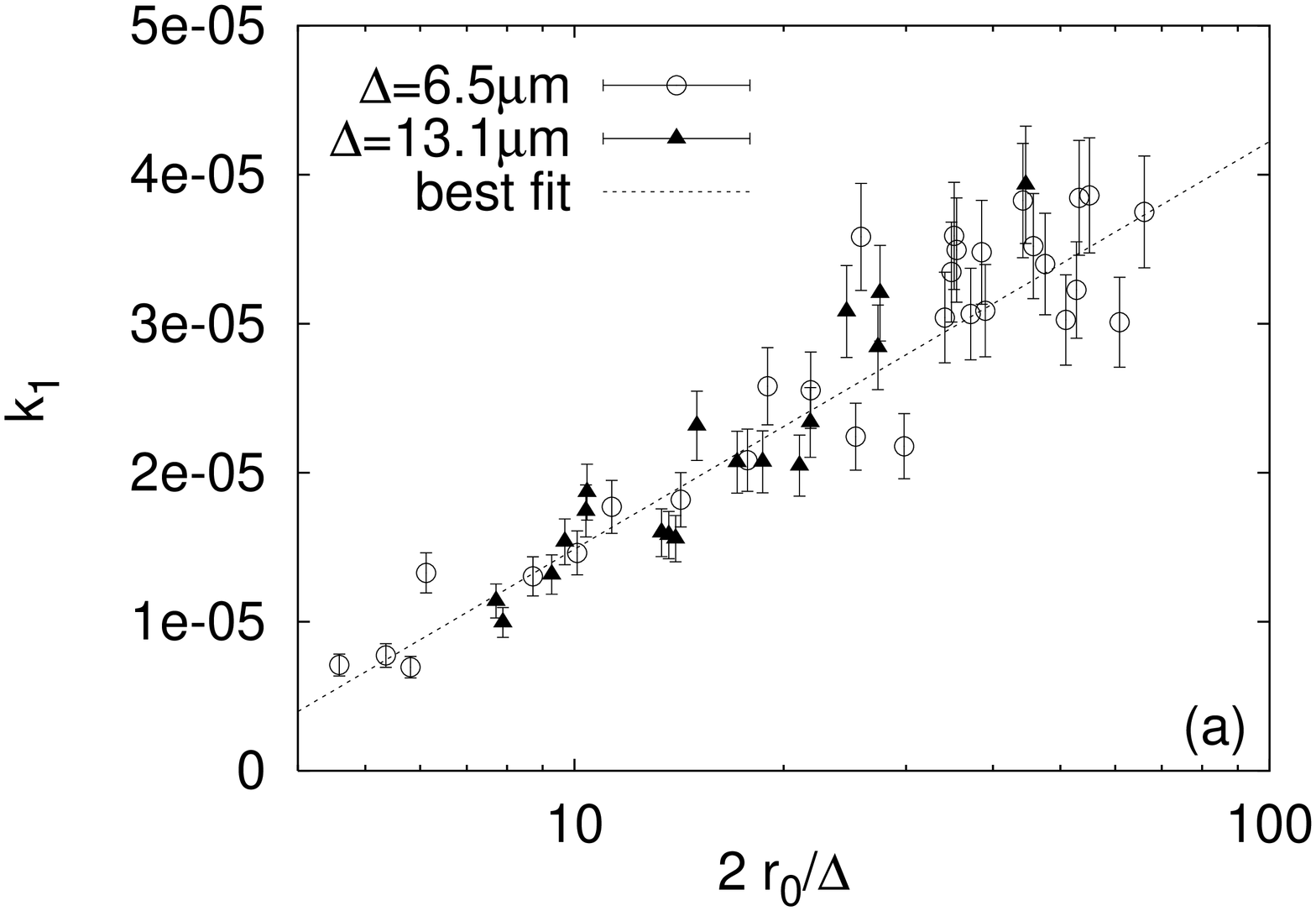,angle=-90,height=4.0in,width=6.0in}
\psfig{figure=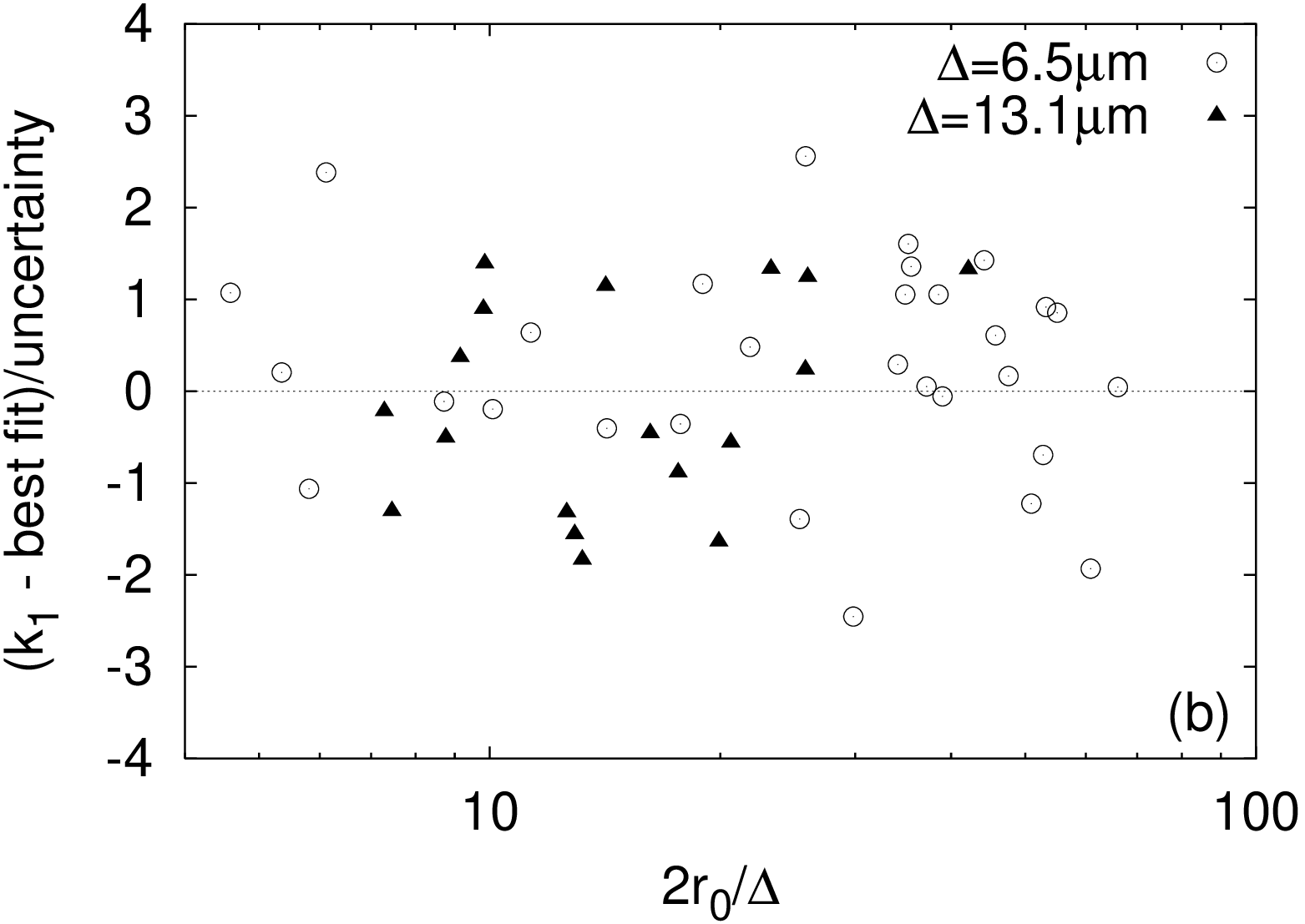,angle=-90,height=4.0in,width=5.45in}
\vspace{0.5cm}
\caption{(a) The plot of $k_1$ vs $1/p= (2r_0/\Delta)$ on a log scale. The
dashed line is the best fit of the data to a straight line. (b) The deviation 
plot of the data from the best fit normalized by uncertainty of each data 
point.}
\label{f5}
\end{figure}

\begin{figure}[tb]
\psfig{figure=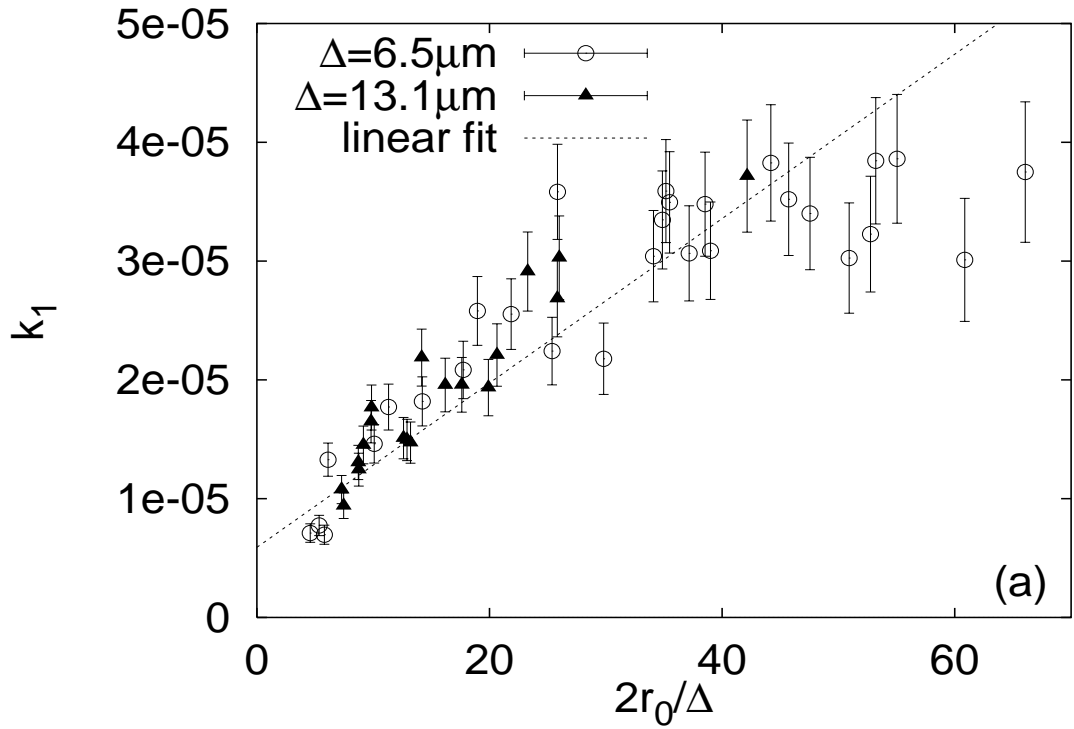,angle=-90,height=4.0in,width=6.0in}
\psfig{figure=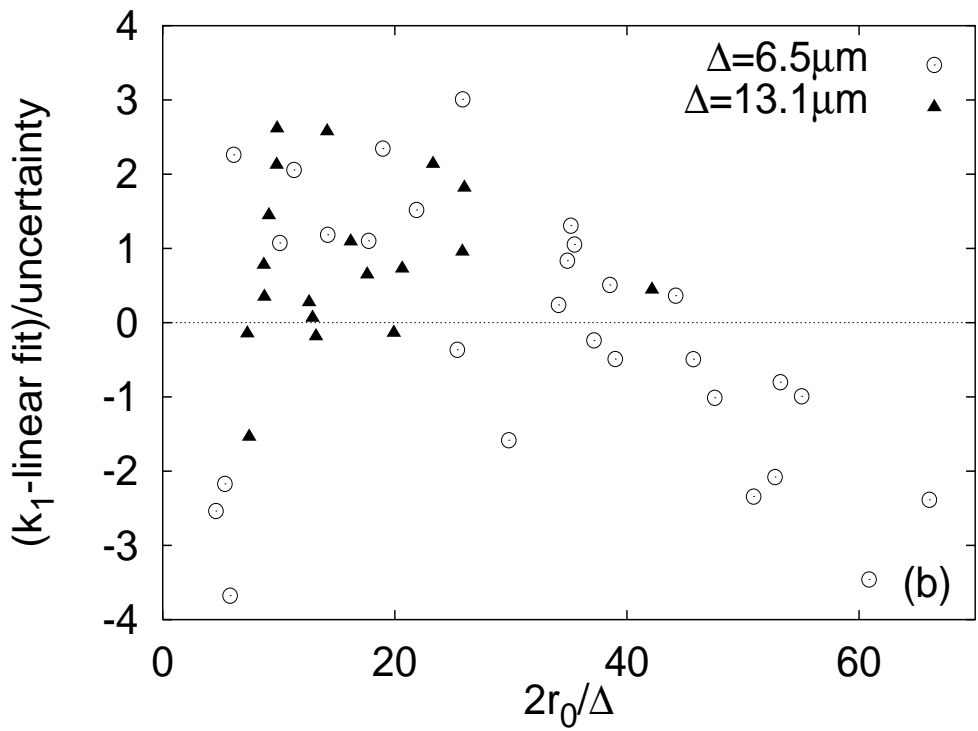,angle=-90,height=4.0in,width=5.45in}
\vspace{0.3cm}
\caption{(a) The plot of $k_1$ vs $1/p=2r_0/\Delta$ and the best fit straight
line. (b) The deviation plot of the data from the best linear fit.}
\label{f6}
\end{figure}

\begin{figure}[tb]
\epsfxsize=6in \epsfbox{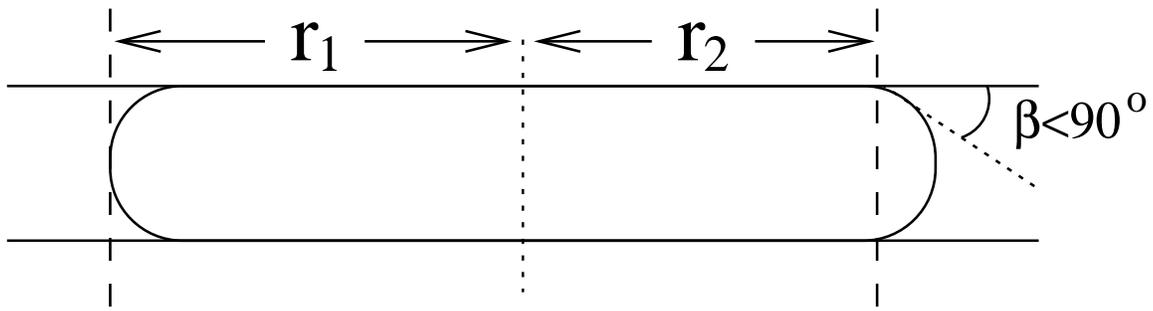}
\caption{A ferrofluid droplet making an acute contact angle with the glass 
plates.}
\label{f7}
\end{figure}
\end{document}